
%
%
%

\input harvmac

\def\np#1#2#3{Nucl. Phys. {\bf B#1} (#2) #3}
\def\pl#1#2#3{Phys. Lett. {\bf #1B} (#2) #3}

\def\physrev#1#2#3{Phys. Rev. {\bf D#1} (#2) #3}
\def\prep#1#2#3{Phys. Rep. {\bf #1} (#2) #3}

\def\ijmp#1#2#3{Int. J. Mod. Phys. {\bf A#1} (#2) #3}

\def\vev#1{\langle#1\rangle}
\def\Tr{{\rm Tr ~}}
\def\tilde{\widetilde}
\def\ti{\tilde i}
\def\tq{\tilde q}
\def\tQ{\tilde Q}
\def\tB{\tilde B}
\def\tb{\tilde b}


\Title{hep-th/9502013, TAUP-2232-95}
{\vbox{\centerline{Remarks on Non-Abelian Duality in N=1}
\centerline{Supersymmetric Gauge Theories}}}
\bigskip
\centerline{Ofer Aharony\foot{This work was supported in part by the
US-Israel Binational Science Foundation, by GIF -- the German-Israeli
Foundation for Scientific Research and by the Clore Scholars Programme.}}
\vglue .5cm
\centerline{School of Physics and Astronomy}
\centerline{Beverly and Raymond Sackler Faculty of Exact Sciences}
\centerline{Tel--Aviv University}
\centerline{Ramat--Aviv, Tel--Aviv 69978, Israel}
\centerline{e-mail: oferah@ccsg.tau.ac.il}

\bigskip\bigskip

\noindent
Recently Seiberg has conjectured a duality symmetry connecting different
theories of the supersymmetric QCD type.
We provide support for this conjecture by
analyzing a flat direction of the theory along which the two dual theories
go over to the same theory in the IR.

\Date{2/95}

\nref\om{C. Montonen and D. Olive, \pl {72}{1977}{117}; P. Goddard,
J. Nuyts, and D. Olive, Nucl. Phys. {\bf B125} (1977) 1}%
\nref\dualstr{A. Sen, hep-th/9402002, \ijmp {9}{1994}{3707}; hep-th/9402032,
\pl {329}{1994}{217}}%
\nref\dualnf{C. Vafa and E. Witten, hep-th/9408074, \np{431}{1994}{3}}%
\nref\swii{N. Seiberg and E. Witten, hep-th/9408099, \np{431}{1994}{484}}%
\nref\swi{N. Seiberg and E. Witten, hep-th/9407087, \np{426}{1994}{19}}%
\nref\klyt{A. Klemm, W. Lerche, S. Yankielowicz and S. Theisen,
``Simple singularities and N=2 supersymmetric Yang--Mills theory",
hep-th/9411048,
CERN-TH.7495/94, LMU-TPW 94/16; P. C. Argyres and A. E. Faraggi,
``The vacuum structure and spectrum of N=2 supersymmetric SU(n) gauge
theory",
hep-th/9411057, IASSNS-HEP-94-94}%
\nref\intse{K. Intriligator and N. Seiberg, hep-th/9408155,
\np{431}{1994}{551}}%
\nref\pwer{N. Seiberg, ``The Power of Holomorphy -- Exact Results in 4D
SUSY Field Theories", to appear in the Proc. of PASCOS 94,
hep-th/9408013, RU-94-64, IASSNS-HEP-94/57}%
\nref\vy{G. Veneziano and S. Yankielowicz, \pl{113}{1982}{231};
T.R. Taylor, G. Veneziano and S. Yankielowicz, \np{218}{1983}{493}}%
\nref\ads{I. Affleck, M. Dine, and N. Seiberg, \np{241}{1984}{493};
\np{256}{1985}{557}}%
\nref\nsvz{V.A. Novikov, M.A. Shifman, A. I.  Vainshtein and V. I.
Zakharov, \np{223}{1983}{445}; \np{229}{1983}{381}; \np{260}{1985}{157}}%
\nref\svholo{M.A. Shifman and A.I Vainshtein, \np{277}{1986}{456};
\np{359}{1991}{571}}%
\nref\cern{D. Amati, K. Konishi, Y. Meurice, G.C. Rossi and G.
Veneziano, \prep{162}{1988}{169} and references therein}%
\nref\nonren{N. Seiberg, hep-ph/9309335, \pl{318}{1993}{469}}%
\nref\nati{N. Seiberg, hep-th/9402044, \physrev{49}{1994}{6857}}%
\nref\dual{N. Seiberg, ``Electric--magnetic duality in supersymmetric
non--abelian gauge theories", hep-th/9411149, RU-94-82, IASSNS-HEP-94/98}%
\nref\iss{K. Intriligator, N. Seiberg and S. Shenker,
hep-ph/9410203, \pl{342}{1995}{152}}%

\newsec{Introduction}

The computation of exact non--perturbative results in quantum field theory
is in general a very difficult problem.
Nevertheless, in the context of supersymmetric
field theories, there has recently been a substantial advance in our ability
to perform such calculations. This advance was
based on two main ideas -- holomorphicity and
strong--weak coupling duality.

Strong--weak coupling duality between electric and magnetic variables
was originally proposed \om\ as a symmetry of certain field theories.
It is now believed that it may in fact have its origin in string
theory \dualstr. This duality (usually called S--duality)
is expected to be an exact
symmetry of scale invariant supersymmetric theories, such as the $N=4$
supersymmetric gauge theory and $N=2$ supersymmetric gauge theories
with certain matter contents \dualnf,\swii. However, it may
also be used in asymptotically free theories which have a dynamically generated
scale $\Lambda$, such as some $N=2$ \refs{\swii-\klyt} and $N=1$ \intse\ gauge
theories. It enables us
to obtain exact non--perturbative information about the Coulomb phase
of these theories, in which the symmetry is broken to an abelian subgroup.
The understanding of these duality transformations in the case of an unbroken
non--abelian gauge symmetry is still far from complete.

The description of supersymmetric field
theories includes several holomorphic objects, whose holomorphicity may be
used (together with the symmetries of the theory and various limits in which
the behavior of the theory is known) to compute them exactly (see \pwer\
for a recent review on this subject). In particular one may obtain by
this method the exact low energy effective lagrangian describing
$SU(N_c)$ supersymmetric
QCD (SQCD) \refs{\vy-\nati} in terms of
gauge--invariant variables, when the number of quark flavors $N_f$ is
at most $N_c+1$. In these cases the theory was found to be in a confining
phase at the origin of moduli space (where the gauge symmetry is unbroken).
However, for $N_f > N_c+1$ it was not possible to find a consistent
IR description
of the theory in terms of gauge--invariant variables.

In a recent paper \dual, Seiberg has conjectured that for $N_f > N_c+1$,
the theory of
SQCD with gauge group $SU(N_c)$ and $N_f$ quark flavors is equivalent
to a theory of SQCD with gauge group $SU(N_f-N_c)$, whose matter content
includes $N_f$ quark flavors
and additional gauge--singlet fields. For $N_f \geq 3N_c$
the original theory is not asymptotically free any more, and
is, therefore, free in the IR, while for
$N_c+2 \leq N_f \leq {3\over 2}N_c$ the dual theory is IR free, so that
it can be regarded
as an appropriate description of the original theory in the IR --
a non--abelian Coulomb phase of QCD. In the
intermediate range ${3\over 2}N_c < N_f < 3N_c$ Seiberg conjectured
that both theories have an IR fixed point, so that the appropriate IR
description of the theory is in terms of some superconformal field theory.
In the case of an $SO(N)$ gauge group, also discussed in \dual, this sort
of duality between theories was shown to be related to the electric--magnetic
duality discussed above. This was possible since in this case the symmetry
may be broken to an abelian subgroup ($SO(2)$) so that there exist
(semi--classically) magnetic monopoles and the electric--magnetic duality
transformation is relatively well understood. However, for an $SU(N_c)$
gauge group with matter in the fundamental representation, the gauge group
cannot be broken (at least semi--classically)
into an abelian subgroup, so that
there are (semi--classically) no magnetic monopoles, and the connection
between the duality of \dual\ and the electric--magnetic duality is, therefore,
still unclear.

The evidence given in \dual\ for the duality conjecture was mostly
kinematical in nature. Seiberg showed that the 't Hooft anomaly conditions
for equating the two theories are satisfied, and that they have the same
flat directions and gauge--invariant observables.  The dynamical evidence
for the duality conjecture relied mainly on the possibility of flowing between
theories with different $N_f$ by giving some quarks a large mass and
integrating them out. Seiberg showed that the duality conjecture
is consistent with this flow, and that by decreasing the number of flavors
to $N_c+1$ or less one regains the known descriptions of SQCD for that
number of flavors, in terms of the quarks or in terms of the
gauge--invariant mesons and baryons. In this letter we wish to provide
more dynamical support for the duality conjecture, by analyzing a flat
direction of the two theories along which both theories become weakly
coupled in the IR, enabling us to directly compare them. This was not
possible in the cases analyzed in \dual\ for which whenever one theory
was weakly coupled the other one was strongly coupled, so that no direct
comparison could be made. In all the cases we will analyze we will find,
in fact,
that both theories go over to the same effective field theory in the IR, so
that obviously they are equivalent there. We consider this to be important
supporting evidence for the duality conjecture.

We begin in section 2 by reviewing the relevant parts of \dual\ which define
the duality transformation. In section 3 we analyze a particular flat direction
of both theories, involving a non--zero vacuum
expectation value of a baryon field.
We find that along this flat direction both theories become free in the IR
with the same massless particle content. In section 4 we add perturbations to
the original theories, and find that after adding them we get,
along the flat direction we
are analyzing, the same interacting IR field theory from both theories. We
end in section 5 with a summary of our results and some speculations.

\newsec{Flat directions in the dual theories}

The duality transformation described in \dual\ connects two different theories
of the supersymmetric QCD type, with generically different gauge groups.
Both theories have enough matter fields in the fundamental representation
so that they are not in a confining phase at the
origin of moduli space. The first theory has only matter fields in the
fundamental representation and no superpotential
(following \dual\ we will call this theory
the ``electric" theory), while the second theory has
additional gauge singlet fields and a non--trivial superpotential (we
will call this the ``magnetic" theory). Here we will only analyze the
case of an $SU(N)$ gauge group.

The ``electric" theory in this case is an $SU(N_c)$ gauge theory
with $N_f$ flavors of
quarks, $Q^i_a$ in the $N_c$ representation and $\tilde Q_{\tilde i}^a$ in
the $\overline N_c$ representation ($i, \tilde i =1, \dots, N_f ; a=1, \dots,
N_c$).  The
anomaly free global symmetry of this theory is (for $N_c > 2$)
\eqn\globsym{SU(N_f)\times SU(N_f) \times U(1)_B \times U(1)_R }
with the quarks transforming as
\eqn\qtransl{\eqalign{
Q &\qquad (N_f,1,1,{N_f-N_c \over N_f}) \cr
\tilde Q & \qquad (1, \overline N_f,-1,{N_f-N_c \over N_f}). \cr}}
The interesting gauge invariant operators which
characterize the moduli space and which we will study
are
\eqn\opsin{\eqalign
{&M^i_{\tilde i}=Q^i Q_{\tilde i} \cr
&B^{[i_1,...,i_{N_c}]}=Q^{i_1}...Q^{i_{N_c}} \cr
&\tilde B_{[\tilde i_1,...,\tilde i_{N_c}]}=\tilde Q_{\tilde
i_1}... Q_{\tilde i_{N_c}} \cr}}
where a summation over color indices in the first definition and an
anti--symmetrization over color indices in the other two definitions are
implied.

For $N_f < N_c+2$ this theory confines, and has an infra--red description
in terms of gauge invariant fields (summarized in \nati).
For $N_f \ge N_c+2$, the only case we will be interested in here,
the quantum moduli space is the same as the
classical one \nati, as can be shown by turning on a tree level
mass term $W_{tree}=\Tr (mM)$ and finding
\eqn\mexpmass{\vev{M^i_{\tilde i}} = \Lambda^{3N_c-N_f \over N_c} (\det
m)^{1 \over N_c} \left({1 \over m}\right)^i_{\tilde i} .}
Then, by studying various limits of $m \rightarrow 0$, all the classically
allowed values of $M$ with $B=\tilde B=0$ can be obtained. Presumably,
by adding other perturbations it can be shown \dual\ that the classically
allowed values with non--zero $B$ and $\tB$ are also in the quantum moduli
space.
Thus, there is no quantum superpotential, and the flat directions of
this theory in which the $D$ terms are zero, are
(up to gauge and global
rotations) of the form
\eqn\flatdiro{Q=\pmatrix { a_1& & & \cr
&a_2& & \cr
& & .  & \cr
& & & a_{N_c}\cr
& & & \cr
& & & \cr} ; \quad
\tilde Q= \pmatrix {
\tilde a_1& &       &         \cr
   &\tilde a_2&     &         \cr
   &   & .   &         \cr
   &   &     & \tilde a_{N_c}\cr
   &   &     &         \cr
   &   &     &         \cr} }
with
\eqn\flatdircon{|a_i|^2 - |\tilde a_i|^2= {\rm independent ~ of} ~ i. }
The gauge invariant description of this moduli space is given in terms of the
observables $M$, $B$ and $\tilde B$ \nati.  Up to global
symmetry transformations they are given by
\eqn\flatmbtb{\eqalign{
M&=\pmatrix { a_1 \tilde a_1& & &  & & &\cr
&a_2 \tilde a_2 & &  & & &\cr
& & .  & & & & \cr
& & & a_{N_c}\tilde a_{N_c} & & &\cr
& & & & & & \cr
& & & & & & \cr} \cr
B^{1,...,N_c}&=a_1a_2...a_{N_c} \cr
\tilde B_{1,...,N_c}&=\tilde a_1 \tilde  a_2...\tilde a_{N_c} \cr}}
with all other components of $M$, $B$ and $\tilde B$ vanishing.  Obviously
the rank of $M$ is at most $N_c$.  If it is less than
$N_c$, either $B=0$ with $\tilde B$ having at most rank one or $\tilde
B=0$ with $B$ having at most rank one.  If the rank of $M$ is equal to
$N_c$, both $B$ and $\tilde B$ have rank one and the product of their
non--zero eigenvalues is the same as the product of
the non--zero eigenvalues of $M$.

The conjectured dual to this theory, the ``magnetic" theory,
is an $SU(N_f-N_c)$ gauge theory,
with $N_f$ flavors of quarks and additional independent singlet fields
corresponding to the mesons of the ``electric" theory.
We will denote the quarks of this theory by
$q_i$ and $\tq^{\ti}$ : $q_i$ transforms as $N_f-N_c$ of the color group and
$\tq^{\ti}$ transforms as $\overline{N_f-N_c}$. The global symmetry group of
this theory is the same as that of the ``electric" theory, given by \globsym.
The quantum numbers
of the quarks in this theory may be determined so that the baryons constructed
from the ``magnetic" quarks have the same quantum numbers as the baryons
constructed from the ``electric" quarks, enabling us to identify them.
This leads to
\eqn\newquaqu{\eqalign{
q \qquad {\rm in} \qquad &(\overline N_f,1,{N_c \over N_f-N_c},
{N_c\over N_f}) \cr
\tilde q \qquad {\rm in} \qquad &(1, N_f,- {N_c \over N_f-N_c},
{N_c\over N_f}) \cr
M \qquad {\rm in} \qquad &(N_f,\overline N_f, 0,
2({N_f-N_c\over N_f})). \cr} }
It is easy to check \dual\
that this assignment of quantum numbers is anomaly free,
and moreover it satisfies the 't Hooft anomaly matching conditions. Thus,
the global
anomalies of this theory
are exactly the same as those of the ``electric" theory.

As it stands, the ``magnetic" theory has an additional gauge--invariant
field which was not present in the ``electric" theory -- the ``magnetic meson"
field $N_i^{\ti} = q_i {\tq}^{\ti}$. To identify the two theories we must
get rid of this field, and this may be done by adding to the action
a superpotential
\eqn\effsup{W= M^i_{\tilde i} q_i \tilde q^{\tilde i}}
(where the color indices are summed).
After adding this term the ``magnetic meson" operator
is redundant -- its coefficient can be absorbed in a shift
of $M$. As we will see, this term is necessary for equating the flat
directions of the two theories, and, as described in \dual, it also
enables us to return to the original theory when performing the duality
transformation twice.

In the ``electric" theory the global symmetry was enhanced
to $SU(2N_f)\times U(1)_R$ when $N_c=2$,
since in that case the quarks and anti--quarks are in the same representation
of the color group. In the ``magnetic" theory we should, therefore, have an
enlarged symmetry group in this case as well. It should, in particular,
relate the mesons and baryons of this theory,
so that its action on the ``magnetic"
quarks will be non--trivial. It is not clear how this
enlarged symmetry of the ``magnetic" theory may be seen directly,
but we will assume its
existence (as part of the duality conjecture).
There is no enhanced symmetry in the ``magnetic" theory when $N_f-N_c=2$,
since it is broken explicitly by the presence of the mesons and by the
superpotential.

In the ``electric" theory we saw that no quantum superpotential was
generated, since all classically allowed values for the meson and
baryon fields could be obtained in the full quantum theory by taking
various limits of massive theories. In the ``magnetic" theory this is
no longer true. This can be seen, for instance, by analyzing the flat
direction
in the classical moduli space in which $M$ has a generic VEV while the
squarks have zero VEV.
Then, all the dual quarks are massive and the low energy ``magnetic" gauge
group leads to gluino condensation.  Working out the $M$ dependence of
the low energy gauge theory, one can find
\dual\ that a superpotential proportional to
\eqn\weffm{{(\det M)^{1 \over N_f-N_c}\over
\Lambda^{3N_c-N_f \over N_f-N_c}}}
is generated.
(We use the conventions of \dual\ for the dimensions of the various
fields, which arrange the powers of the QCD scale
$\Lambda$ according to the dimensions of
the fields at the UV fixed point of the ``electric" theory, and add additional
powers of $\Lambda$ when looking at other limits in which the fields have
different dimensions).
By adding to the action a small
mass term, $\Tr(mM)$, and requiring that we get the
expectation values \mexpmass, we find that the quantum effective
superpotential of the ``magnetic" theory along this flat direction is in fact
\eqn\weff{W_{eff} = M^i_{\ti} q_i {\tq}^{\ti} - (N_f - N_c) {(\det M)^{1 \over
N_f-N_c} \over \Lambda^{3N_c-N_f \over N_f-N_c}}.}
We will assume that this is the exact superpotential along the flat directions
for which the gauge group is confined, or equivalently completely Higgsed
(see \intse\ for a discussion of how different effective superpotentials
can arise in different phases of the theory). These are the flat directions
we will be interested in here. The most general superpotential respecting the
symmetries looks like the first term in \weff\ times a general function of the
ratio of the two terms, but the choice above
seems to be the only one for which the
relevant flat directions in the ``magnetic" theory are indeed equivalent to
those of the ``electric" theory.
It is interesting to note that the additional quantum superpotential
becomes irrelevant in the IR exactly when $N_f < {3\over 2} N_c$
(since the power of the meson superfield appearing is $N_f \over {N_f-N_c}$
which is larger than $3$ in this case). Thus,
in this case the ``magnetic" theory indeed becomes free in the IR.

The flat directions of the ``magnetic" theory may now be determined by
requiring that the $D$ terms and the $F$ terms coming from $W_{eff}$ all
vanish. It can easily be seen that if either $q$ or $\tq$ is zero, an
expectation value of $M$ of rank less than $N_c$ is a flat direction
of this theory. The $D$ term in this case forces $q$ (or $\tq$) to have
$N_f-N_c$ equal eigenvalues. If both $q$ and $\tq$ are non--zero, the
$D$ and $F$ terms force both of them to be of rank $N_f-N_c$ (so that
$B$ and $\tilde B$ are non--zero),
with the matrix $\tq q$ having
$N_f-N_c$ equal non--zero eigenvalues.
$M$ then has to be of rank $N_c$, with the
product of the eigenvalues of $M$ equaling (up to an appropriate power of
$\Lambda$) the product of the non--zero
elements of $B$ and $\tilde B$.

We have thus found exactly the same flat directions in both theories when
expressed in terms of the gauge--invariant variables. In both theories
there is one flat direction in which the meson $M$ has a rank less than $N_c$
and $B$ and $\tB$ are both
zero or at most one of them has one non--zero eigenvalue. Another flat
direction is the one in which $B$ and
$\tB$ both have one non--zero eigenvalue, in which case $M$ is forced
to be of rank $N_c$ with the product of its non--zero eigenvalues equal to the
product of the non--zero eigenvalues of $B$ and $\tB$.

\newsec{The baryonic flat direction}

Most of the flat directions described above were analyzed in \dual\ for
both theories, and it was shown that by going along them and integrating
out the fields that become massive, the IR description of the two theories
becomes that of another pair of dual theories. An interesting flat
direction which was not analyzed in \dual\ is the direction in which
the baryon $B$ gets a VEV while the anti--baryon $\tB$ does not. In
this case, which we will call the baryonic flat direction,
we will show that the two theories in fact go over to exactly
the same theory in the IR. We consider this to be important support
for the conjecture that the two theories are in fact the same
in the IR.

Let us begin with the simplest case, in which the meson expectation
value is zero.
In the ``electric" theory this corresponds to squark
expectation values of the form (up to gauge and global rotations)
\eqn\flatdire{Q=\pmatrix { a& & & \cr
&a& & \cr
& & .  & \cr
& & & a\cr
& & & \cr
& & & \cr} ; \quad
\tilde Q= 0 }
and in the ``magnetic" theory this corresponds to
\eqn\flatdirm{q=\pmatrix {
& & & \cr
& & & \cr
A& & & \cr
&A& & \cr
& & .  & \cr
& & & A\cr}
 ; \quad
\tilde q= 0 ; \quad M = 0}
where the baryon  VEV in the ``electric" theory equals $a^{N_c}$, and
the baryon VEV in the ``magnetic" theory (related to the ``electric"
VEV by appropriate powers of $\Lambda$) equals $A^{N_f-N_c}$.
In both cases the gauge symmetry is completely broken, so that both theories
are in a Higgs phase. This is
unlike the cases analyzed in \dual\ for which one theory
was in a Higgs phase when the other was in a confining phase, although the
two phases are of course indistinguishable in this case since we have
matter only in the fundamental representation of the gauge group.
Note that the baryon composed of the first $N_c$ quarks in the ``electric"
theory is equivalent to the baryon composed of the last $N_f-N_c$ quarks
in the ``magnetic" theory.

In the ``electric"
theory, the $SU(N_f)$ flavor symmetry acting on the quarks $Q$ is broken
to $SU(N_c)\times SU(N_f-N_c) \times U(1)_F$. The two original $U(1)$
symmetries are explicitly broken by the VEV, but a combination of each
one of them with the $U(1)_F$ from the flavor symmetry remains unbroken.
A diagonal sub--group of the product of the $SU(N_c)$ color symmetry and
the $SU(N_c)$ coming from the flavor group also remains unbroken, so that
the total global symmetry of the theory is
\eqn\globsymb{SU(N_c) \times SU(N_f-N_c)
\times SU(N_f) \times U(1)_{\hat B} \times U(1)_{\hat R}.}
Of the first $N_c$ quark
flavors (each one consisting of $N_c$ chiral superfields),
$N_c^2-1$ chiral superfields join the gauge bosons by the
Super--Higgs mechanism to become $N_c^2-1$ massive vector multiplets of
mass $g_e |a|$ (where $g_e$ is the coupling constant of the ``electric"
theory). The remaining combination is a massless singlet
labeling the flat direction which will not play a role in the subsequent
discussions. The other quarks and anti--quarks all remain massless, and
their quantum numbers under the new global symmetry can easily be found
to be
\eqn\qtranse{\eqalign{
Q &\qquad (N_c,N_f-N_c,1,{N_f \over N_f-N_c},1) \cr
\tilde Q & \qquad (\overline N_c,1, \overline N_f,-1,{N_f-N_c \over N_f}).
\cr}}
The massless $Q$ quarks
(and one real scalar from the massless singlet mentioned above)
may be interpreted as Goldstone bosons for the breaking of the global flavor
symmetry.
Integrating out the massive fields leaves us in this case with a free theory
with this field content.

The ``magnetic" theory along this flat direction behaves in a similar
fashion. The $SU(N_f)$ flavor symmetry acting on the quarks breaks here
as well to $SU(N_f-N_c)\times SU(N_c) \times U(1)_F$, with a diagonal
subgroup of the $SU(N_f-N_c)$ color symmetry and the $SU(N_f-N_c)$ symmetry
from the breaking of the flavor group remaining a global symmetry of the
theory. As above, $(N_f-N_c)^2-1$ quarks join the gauge bosons to become
massive vector superfields of mass $g_m |A|$, while another combination
of the first $N_f-N_c$ quarks is massless and labels the flat direction.
However, in this case the superpotential also gives a mass to some of the
other fields. All the anti--quark fields $\tq$ get a mass $|A|$, as do
the mesons $M^i_{\ti}$ for $i=N_c+1,\dots,N_f$. The fields that remain massless
are, thus, $N_c$ flavors of quarks, and $N_c N_f$ meson fields.
The global symmetries that remain are exactly the same as those we found
in \globsymb\ for the ``electric" theory, and again it is easy to compute the
quantum numbers of the remaining massless fields, which are
\eqn\qtransm{\eqalign{
q &\qquad (\overline N_c,N_f-N_c,1,{N_f \over N_f-N_c},1) \cr
M &\qquad (N_c,1, \overline N_f,-1,{N_f-N_c \over N_f}). \cr}}
As in the ``electric" theory, the massless fields $q$ may be interpreted
as Goldstone bosons for the global symmetry breaking.

Up to conjugation of the $SU(N_c)$ subgroup of the global symmetry group
(which is obviously a matter of convention), these are exactly the same quantum
numbers as those we found above for the massless fields $Q,\tQ$ in the
``electric"
theory. Here too we can integrate out the massive fields, and we find
that the remaining superpotential for the massless fields is zero, so
that the theory is free in the IR. The two dual theories thus
lead to exactly the same theory in the IR along the baryonic flat direction,
providing support for the conjecture that the theories are indeed the same
in the IR. Of course, this is a trivial case since both theories are free
in the IR. Later we will see examples where the two dual theories go over
to the same interacting field theory in the IR.

Once again, special notice should be given to the $N_c=2$ case, in which
the ``electric" theory has a larger symmetry. Along the flat direction
we are analyzing, the global symmetry in this case is $SU(N_c)\times
SU(2N_f-N_c) \times U(1)_{\hat R}$ and all massless quarks sit in the same
fundamental representation of $SU(2N_f-N_c)$. For the duality to hold,
we need a similar symmetry connecting the massless fields $q$ and $M$
in the ``magnetic" theory. We will argue that the symmetry we need
can in fact be derived from
the symmetry connecting the mesons and the baryons in the ``magnetic"
theory (which we assume exists in the IR). This is the case since
when we replace the
heavy quarks by their VEVs (when integrating them out), we find that the
baryon containing $N_f-N_c-1$ of the last $N_f-N_c$ (massive) quark flavors,
and one of the remaining $N_c$ massless flavors,
equals $A^{N_f-N_c-1}$ times a
component of the massless quark flavor. This component has its
color index equal
(up to a shift by $N_c$) to the
flavor index missing (among the last $N_f-N_c$ flavor indices) from the
definition of the baryon. Thus, the massless quarks in this case
are equivalent to baryons
(as found in \dual\ also for other cases) so that the additional symmetry
required follows straightforwardly from the symmetry between the mesons and the
baryons.

The above results can easily be generalized to the case in which the
meson matrix also gets a VEV of rank $k$ for $k < N_c$ (with the same
baryon VEVs). For example, if we have a meson VEV with $k$ non--zero
eigenvalues which are all equal, the global symmetry breaks in both
theories into
\eqn\globsymbm{SU(k) \times SU(N_c-k) \times
SU(N_f-k) \times SU(N_f-N_c) \times U(1)
\times U(1)_B \times U(1)_R}
and again we can find in the IR a free field
theory with the same massless particle content originating from both theories.

An analysis similar to the one we did in this chapter may also be performed
for the case of $N_f=N_c+1$, whose exact superpotential was given in \nati.
Indeed, if we identify the baryon field used there with the dual quark
(up to an appropriate scale factor),
the superpotential given in \nati\ is exactly the same as \weff. Thus, the
analysis for that case is exactly the same as the one we performed here,
so that also there the two descriptions go over to the same theory in the IR
along the baryonic flat direction.

\newsec{Perturbations along the baryonic flat direction}

In order to verify that the duality conjecture holds, one should
in principle compute correlation functions in both theories and see that they
give the same result. For example, since the global symmetry is the same in
both theories, one can always try to compare correlation functions of global
symmetry currents.
Unfortunately, since the gauge coupling in one theory
gets stronger when that of the other theory gets weaker,
it seems impossible to perform such computations
perturbatively in both theories as long as the
gauge symmetry is unbroken.

However, we have seen that there is a flat
direction along which the two theories both go into the same theory in the IR,
so that along this flat direction we can directly compare the IR correlation
functions of the two theories. These are, however, trivial
in this case since both theories
are free below the symmetry breaking scale. It would be more interesting
if we could find a phase in which both theories would go over to a non--trivial
weakly coupled theory in the IR. Then we could compare non--trivial
correlation functions between the two theories. We will do this by adding to
the action perturbations by various gauge invariant superfields, which do
not destroy the baryonic flat direction. We will find in all
such cases that the resulting effective action in the
IR is the same in both theories,
so that obviously all correlation functions are the same.
For simplicity we will only analyze in this section the case in which
the meson VEV is zero.

The gauge invariant objects we can perturb the action by are essentially
the meson and baryon operators. The simplest perturbation is obviously by
a meson operator, and if we choose it to be a meson operator which does
not include any of the first $N_c$ quark flavors, it will not destroy
the baryonic flat direction. Let us choose \dual\ a perturbation of the
form $W=mM^{N_f}_{N_f}$.
Note that the dimension of $m$ in the UV of the ``electric" and ``magnetic"
theories is not the same. We will denote the constant $m$ in the ``electric"
theory by $m_e$, and in the ``magnetic" theory by $m_m$, and we will find
below what the relation between them must be.

In the ``electric" theory, all this perturbation does is to give
a mass $m_e$ to the last quark and anti--quark flavors, which were so far
massless.
If $m_e$ is much smaller than the symmetry breaking scale $a$, we can
integrate out the same massive fields as above, and remain with the same
field content as in \qtranse\ but with a mass $m_e$ for one quark and one
anti--quark flavor. In the ``magnetic" theory,
however, the addition of this term
changes the $F$-term equation of the relevant meson (which is massive there)
to $q_{N_f} \tq^{N_f} +
m_m = 0$. Thus, if we wish to keep the same flat direction as before,
we must have $\vev{\tq^{N_f}_{N_f-N_c}} = -{m_m\over A}$. In fact
$\vev{q_{N_f}^{N_f
-N_c}}$ will in this case be corrected by terms
of order ${m_m^2}\over {A^3}$ to keep the $D$-terms
zero, but we will assume $m_m << A^2$ and ignore these terms (the opposite
case of large $m_m$, for which the fields of mass $m_m$ are integrated out, was
analyzed in \dual).
Through the superpotential, this leads to a mass $|{m_m\over A}|$ for the
quarks $q_i^{N_f-N_c}$ of all flavors $i$, and for the mesons $M_{N_f}^i$.
After integrating out the fields whose mass is of order $A$,
we remain with a free theory, with
a mass $|{m_m\over A}|$ for the
$(N_f-N_c)$-th color component of the quarks which were
massless before, and for the mesons $M_{N_f}^i$ for $i=1,\dots,N_c$ which
were massless before. These are exactly the particles that we identified
above (in the discussion of the free theory)
with the last flavor of quark and anti--quark in the ``electric" theory. Thus,
we find that (up to a factor of $A$ which we means we should identify
$m_m=Am_e$) the
two dual theories again go over to the same theory in the IR. Again it is a
free theory but this time with a non--trivial
(and equal) mass spectrum. Analogous
results can also be obtained
by analyzing more complicated mesonic perturbations,
as long as they leave the baryonic flat direction flat.

The next perturbation one can think of is a perturbation by the baryon
operators
$B$ and $\tB$. These terms are non--renormalizable in the ``electric" theory if
$N_c > 3$ and in the ``magnetic" theory if $N_f-N_c > 3$. However,
since we are only
looking at the IR effective action, we can add them even in those cases (as
done
for instance in \iss), assuming that they can be derived
from some renormalizable
interaction at higher energies. Let us begin by analyzing the perturbation by
one baryon operator, which is in the ``electric" theory
\eqn\pertb{W = b_e
B^{i_1,\dots,i_{N_c}} = b_e \det (Q^i_a)_{i=i_1,\dots,i_{N_c},a=1,\dots,N_c}.}
This perturbation does not destroy the baryonic flat direction as long
as at least two of the quark flavor indices
involved are greater than $N_c$. We
will analyze here the simplest case in which all flavors involved are greater
than $N_c$ and
$N_f = 2N_c$. In this case all the
quarks involved remain massless along the baryonic flat direction. We
are left,
after integrating over the massive fields, with the same superpotential
\pertb\
(which leads to a non--trivial interaction near the IR), which is simply
the determinant of the matrix of massless quark fields.  The analogous
interaction in the ``magnetic" theory is the same, with the baryon operator
equal to the determinant of the first $N_c=N_f-N_c$ quark flavors.
Once again all these quarks remain massless along the baryonic flat direction.
We remain, after integration over the massive fields,
with a superpotential which includes only the perturbation, which is again
the determinant of the matrix of massless quarks. This is the first example
in which we find that the two
dual theories go over to the same {\it interacting}
field theory in the IR limit.
Other baryon perturbations, for general values of $N_f$,
work in a similar fashion, leading in all cases to
the same superpotential in both theories
after integrating out the massive fields (which for this type of perturbation
is generally
equivalent to replacing the massive quarks by their VEVs).

A more interesting case is adding a perturbation by the anti--baryon
operator $\tilde B$. The baryonic flat direction survives this
type of perturbation for any value of $N_f,N_c$ and for any choice of
anti--quark flavors composing the anti--baryon. For definiteness let us take
the
perturbation in the ``electric" theory to be
\eqn\perttbe{W_e = \tb_e \tB = \tb_e \det(
\tQ^a_{\ti})_{a,\ti=1,\dots,N_c}}
and in the ``magnetic" theory it is
then
\eqn\perttbm{W_m=\tb_m
\det(\tq^{\ti}_a)_{a=1,\dots,N_f-N_c,\ti=N_c+1,\dots,N_f}.}
In the ``electric" theory this perturbation is similar to the previous
case. It does not affect the baryonic flat direction,
and remains after integrating
out the massive fields, since all the fields involved are massless. In the
``magnetic" theory, however, the fields involved are all massive and one
should carefully integrate out the massive fields using a superpotential
that is the sum of $W_{eff}$ from \weff\ and $W_m$. Performing this
calculation leads to an effective potential of the form
\eqn\wefftb{W_{eff} = \tb_m {\Lambda^{-(3N_c-N_f)} \over A^{N_f-N_c}}
\det(M^i_{\ti})_{i,\ti=1,\dots,N_c}.}
Up to
scale factors arising from transforming the relevant operators to the
``magnetic" theory, this is exactly the same interaction as
we found in the ``electric" theory, since the mesons involved were
identified (in the IR where we have a free theory)
with the anti--quarks involved
in the ``electric" interaction. This is another non--trivial example of
obtaining the same IR theory when starting from the two dual theories.

In a similar fashion one can analyze more complicated perturbations, like
powers of meson operators. In all cases that were checked one obtains
the same interacting IR theory when starting from the two dual theories and
integrating out the massive fields along the baryonic flat direction.
Namely, the interaction in the
``electric" theory in terms of the massless quarks
and anti--quarks turns out to be
the same as the interaction in the ``magnetic" theory
in terms of the massless quarks and mesons. For this to hold the ``electric"
and
``magnetic" quarks and the ``electric" anti--quarks and ``magnetic" mesons
are identified as in the free theory we analyzed in the previous section.

The analysis of this chapter may also be generalized to the case $N_f=N_c+1$.
There is no baryon perturbation in that case, since it always ruins the flat
direction we are checking. However, the meson and anti--baryon perturbations
work
there exactly the same as they do here, giving once again
the same interacting IR theory starting
from the original two dual descriptions.

\newsec{Summary and Conclusions}

Establishing the duality conjecture of \dual\ in a convincing manner
requires the comparison of physical quantities between the ``electric"
and ``magnetic" theories. Such quantities could be, for instance, correlators
of the global symmetry currents, or some critical exponents like the derivative
of the beta function at the IR fixed point\foot{We would like
to thank M. Peskin for discussions on this subject.}. Unfortunately this
sort of comparison is hard to achieve, since usually at least one of the
theories is strongly coupled. In this letter we provided support for the
duality conjecture by studying a particular flat direction along which
both theories become weakly coupled in the IR. We found that, along this
flat direction, both theories give rise to the same effective theory in
the IR. Hopefully our calculations, and perhaps similar calculations
which may be performed for other gauge groups,
will assist in shedding some light on the
origin of this duality symmetry, and on its relation to the electric--magnetic
duality, which are still not clear.

The duality symmetry in \dual\ is presented as an equality between the
IR descriptions of two theories, and not as an exact equivalence between
the two theories at all energy scales.
It is an intriguing open question to compare the theories also in the UV.
Naively, in the range ${3\over 2}N_c < N_f < 3N_c$ for which both
theories are asymptotically free, we would expect the two theories to
be different in the UV, as would be revealed (upon gauging part of
the global symmetry) by experiments such
as $e^+ e^-$ annihilation and deep inelastic scattering. In particular the
value of $R$ in the two theories is expected to be different since the
quark content of the theories is different.
It seems to us, however, that one should be more cautious
in using arguments such as this.
When the
gauge coupling of the ``magnetic" theory becomes weak in the UV, the
superpotential becomes strong (since it is like a $\Phi^3$ coupling in
the WZ model when the gauge coupling can be neglected), so that it does
not seem possible to actually perform any trustworthy perturbative
calculations in the UV of the
``magnetic" theory. Perhaps the superpotential only becomes strong at
a scale much higher than $\Lambda$,
in which case perturbative calculations would
be trustworthy at energies
between the two scales and would rule out the equivalence
between the two theories. However, as long as the behavior of the coupling
in the superpotential as a function of the scale is not well understood,
it does not seem possible to rule out the possibility that the two theories
will somehow conspire to give the same results in the UV as well.
Clearly a better understanding of this issue is important in determining
whether the theories could be the same in the UV or not.

Another issue is that, assuming that for ${3\over 2}N_c < N_f < 3N_c$
the beta function has exactly one zero, the theory exists (at least formally)
in one of three regimes. The gauge coupling is either
exactly at the critical coupling,
or it is below it and becomes weakly coupled in the UV, or it is above
it and becomes strongly coupled in the UV. Since the duality transformation
is supposed to
interchange strong and weak coupling, it may very well be
that the theory in
the asymptotically free regime would be equivalent by duality to
a theory in the non--asymptotically free regime. If this is true, a
comparison between the theories when both are asymptotically free is not
relevant. Note that both theories still must have the same behavior in the
IR. A similar argument can of course be made when the beta function
has more than one zero.

Obviously, much more work is needed in order to understand the origin and
nature of this conjectured duality, and whether it holds only for the IR
description of the theory or at higher energies as well.

\bigskip

\centerline{\bf Acknowledgements}

We would like to thank B. Blok, O. Ganor and N. Seiberg for useful
discussions, and especially J. Sonnenschein and S. Yankielowicz for
many useful discussions and for a critical reading of this manuscript.

\listrefs

\end